\documentclass[twocolumn,showpacs,preprintnumbers,amsmath,amssymb]{revtex4}

\usepackage{graphicx}
\usepackage{dcolumn}
\usepackage{bm}
\begin{document}

\title{Hierarchical Spatial Organization of Geographical Networks}

\author{Bruno A. N. Traven\c{c}olo}
\author{Luciano da F. Costa}%
\affiliation{%
Cybernetic Vision Research Group, GII-IFSC, Universidade de S\~{a}o
Paulo, S\~{a}o Carlos, SP, Caixa Postal 369, 13560-970, Brazil,
luciano@if.sc.usp.br.
}%

\date{\today}

\begin{abstract}
In this work we propose the use of a hirarchical extension of the
polygonality index as a means to characterize and model geographical
networks: each node is associated with the spatial position of the
nodes, while the edges of the network are defined by progressive
connectivity adjacencies. Through the analysis of such networks, while
relating its topological and geometrical properties, it is possible to
obtain important indications about the development dynamics of the
networks under analysis.  The potential of the methodology is
illustrated with respect to synthetic geographical networks.
\end{abstract}

\maketitle

Geographical complex networks have been used to model biological
systems - e.g. bone structure~\cite{Costa2006b} and mammalian cortical
areas~\cite{Costa2007b}. In these system, the physical proximity and
communication between neighboring elements are critical to guarantee
proper development and biological function. Another important property
of these systems is its spatial organization. It is known that many
biological systems depend on the proper adjacency of cells for correct
development - e.g. cell communication, distribution of cells in
retina, kidney structures, among many others.

Despite the large set of tools and measurements used to characterize
the networks that underly these systems~\cite{Costa2007}, relatively
little attention has been given to the spatial organization of the
nodes. In order to address this issue, we propose the use of the
polygonality index~\cite{Costa2006} -- a robust measurement able to
quantify the spatial order of system of points -- over the nodes of a
geographical network.

\begin{figure}
  \includegraphics[width=5cm]{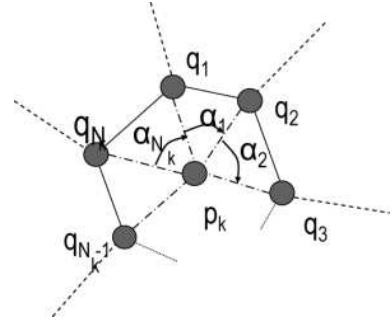}\\
  \caption{{\bf Quantifying the spatial organization around a
  node}. All the angles $\alpha_i$ formed between successive neighbors
  ($q_i$) of the node $p_k$ are used to compute the polygonality index
  (see Eq. 1).}\label{fig:ang}
\end{figure}

The polygonality index assigns to each node of the network a value
that indicates the amount of the organization around the node with
respect to its neighbors. This value is computed based on the angle
$\alpha_i$ formed between successive neighbors of a node, as
exemplified in Fig.~\ref{fig:ang} and expressed by the following
equation:

\begin{equation}
h_k = \frac{1}{\sum_{i=1}^{N_k}|\alpha_i-\beta| + 1}
\end{equation}

where $k$ is the node under analysis, $N_k$ is the number of neighbors
of the node $k$ and $\beta_k$ is a parameter whose value can be fixed
-- to characterize specific spatial arrangements -- or can vary
accordingly to the number of the neighbors of the nodes, i.e, $\beta =
2\pi/N_k$.  A fixed value of $\beta$ allows one to identify whether
the spatial position of the adjacent nodes obeys a specific
arrangement (hexagonal, orthogonal, etc.). For example, if $\beta =
\pi/3$, it is possible to quantify how much the nodes of network
differ from an hexagonal arrangement. When the value of $\beta$
depends on the number of the neighbors, the polygonality index
indicates if the angles between neighbors are equally distributed.
In both cases, the polygonality index varies between 0 (total lack
of spatial order) and 1 (fully organized system).

\begin{figure*}
  \includegraphics[width=16cm]{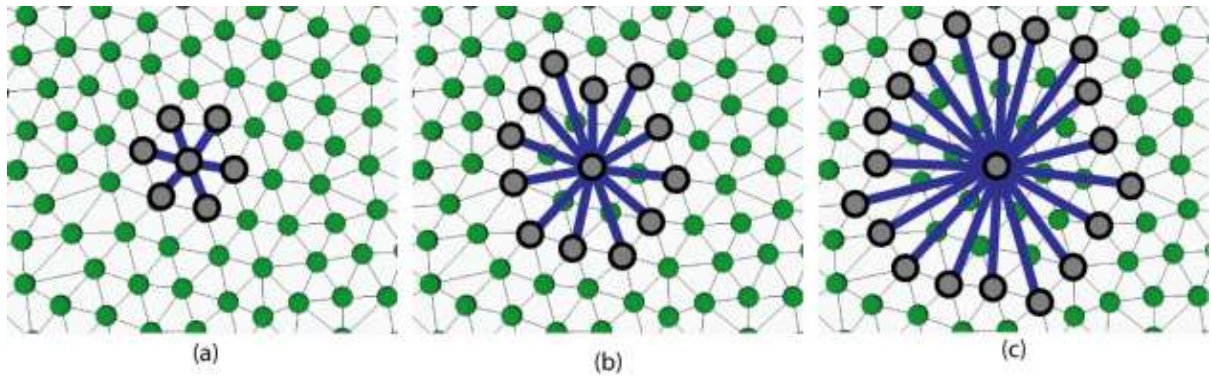}\\
  \caption{{\bf Hierarchical neighbors used to estimate the spatial
  organization}. Hierarchical neighbors 1 (a), 2 (b) and 3 (c). The
  virtual edges (in blue) between the central node and its
  hierarchical neighbors are used to estimate the angles $\alpha_i$
  for polygonality index computation (see Eq. 1) .}\label{fig:hie}
\end{figure*}

In this article we extend the concept of polygonality in order to
allow the quantification of the spatial order in a complex network
including several hierarchical levels, allowing the quantification of
the spatial organization along a wider neighborhood around each
node. Figure~\ref{fig:hie} illustrates the computation of such a
polygonality index for three different hierarchical levels of a
node. Note that {\it virtual edges}\cite{Costa2006c} are established
between the central node and its hierarchical neighbors. The angles
between successive virutal edges are taken into account to compute the
polygonality index, as expressed in Eq. 1.

\begin{figure*}
  \includegraphics[width=16cm]{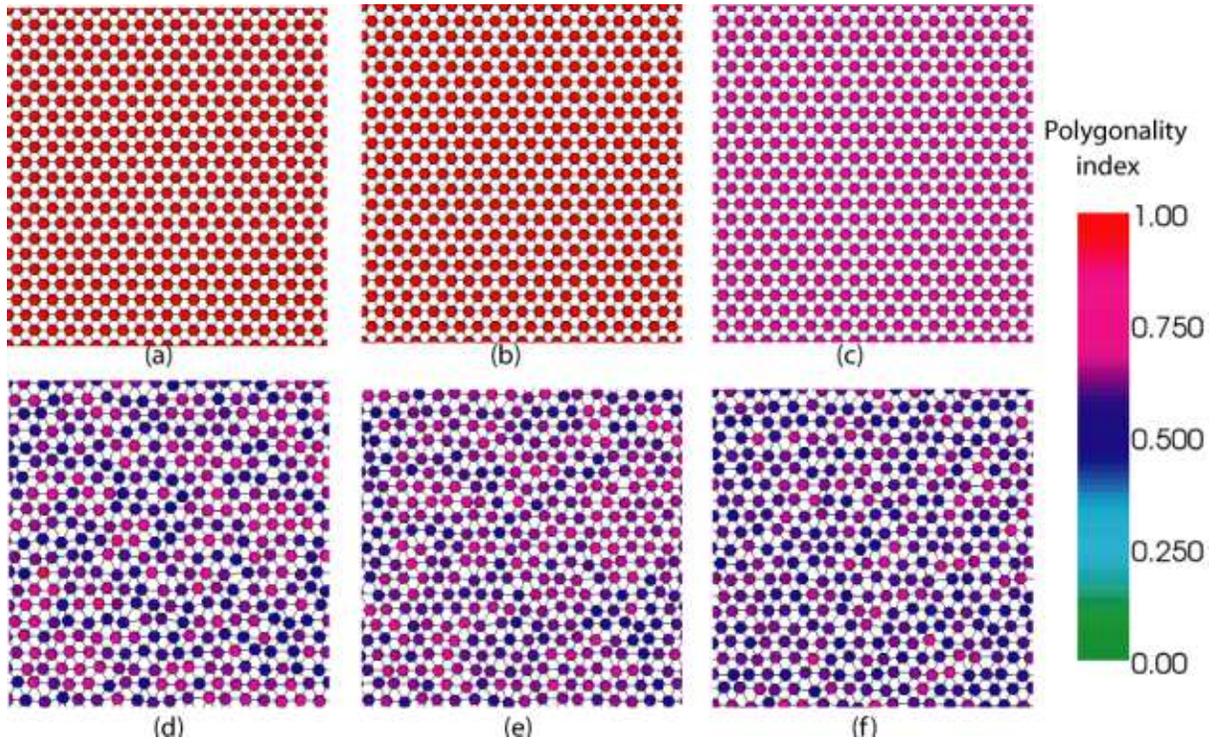}\\
  \caption{Hierarchical polygonality index for a
  perfect hexagonal lattice (a-c) and for a perturbed hexagonal
  lattice (d-e). The hierarchial level ($R$), the mean ($\mu$) and standard deviation ($\sigma$)
  values of the polygonality index are: (a) $R=1$,
  $\mu = 1$,  $\sigma = 0$; (b) $R=2$,$\mu = 1$,  $\sigma = 0$;(c) $R=3$, $\mu = 0.73$, $\sigma = 0$;
  (d) $R=1$, $\mu = 0.63$, $\sigma = 0.07$; (e) $R=2$, $\mu = 0.64$, $\sigma =
  0.05$; (f) $R=3$, $\mu = 0.61$, $\sigma =  0.05$.}\label{fig:examples}
\end{figure*}

Figure~\ref{fig:examples} shows six networks and respective
polygonality indices. In (a-c) the networks are perfectly arranged in
hexagonal lattices, while in (d-f) the networks originated from a
hexagonal lattice, but the positions of their nodes were perturbed in
order to reduce the overall order. In (a) and (d) the first hierarchal
level was considered to estimate the polygonality index, while in (b)
and (e) the second level was used. The results in (c) and (f)consider
the third level. In all these images the nodes were colored
accordingly to their polygonality index, as expressed in the
color-bar.

Analyzing the values of the polygonality index in
Figure~\ref{fig:examples}, it is possible to note that in (a), as the
nodes and its neighbors are arranged in a hexagonal way, the
polygonality index for all nodes is the maximum possible. In (b), the
angles defined between successive neighbors of the nodes are constant
and equal $\beta = \pi/6$. As consequence, the polygonality index is
also at its maximum value. On the other hand, in (c) the angles
between successive neighbors of the third hierarchy are no longer
constant, but the polygonality index is the same for all nodes. In the
case of the perturbed lattices (Figure~\ref{fig:examples}(d-f)), the
polygonality index for all hierarchies are, as expected, lower than
the regular lattices.

All in all, the presented measurements allowed a comprehensive
characterization of the spatial organization of geographical complex
networks. They can also be compared with several other measurements
related to the dynamics and connectivity of the network, paving the
way to obtaining new insights about the topological and geometrical
properties of these networks.

\section*{Acknowledgments}

Luciano da F. Costa is grateful to CNPq (301303/06-1) and FAPESP
(05/00587-5). Bruno A. N. Traven\c{c}olo thanks to FAPESP for financial
support (03/13072-8).

\bibliography{cagliari_bruno}

\begin{thebibliography}{5}
\expandafter\ifx\csname natexlab\endcsname\relax\def\natexlab#1{#1}\fi
\expandafter\ifx\csname bibnamefont\endcsname\relax
  \def\bibnamefont#1{#1}\fi
\expandafter\ifx\csname bibfnamefont\endcsname\relax
  \def\bibfnamefont#1{#1}\fi
\expandafter\ifx\csname citenamefont\endcsname\relax
  \def\citenamefont#1{#1}\fi
\expandafter\ifx\csname url\endcsname\relax
  \def\url#1{\texttt{#1}}\fi
\expandafter\ifx\csname urlprefix\endcsname\relax\def\urlprefix{URL }\fi
\providecommand{\bibinfo}[2]{#2}
\providecommand{\eprint}[2][]{\url{#2}}

\bibitem[{\citenamefont{da~F.~Costa
  et~al.}(2006{\natexlab{a}})\citenamefont{da~F.~Costa, Viana, and
  Beletti}}]{Costa2006b}
\bibinfo{author}{\bibfnamefont{L.}~\bibnamefont{da~F.~Costa}},
  \bibinfo{author}{\bibfnamefont{M.}~\bibnamefont{Viana}}, \bibnamefont{and}
  \bibinfo{author}{\bibfnamefont{M.}~\bibnamefont{Beletti}},
  \bibinfo{journal}{Applied Physics Letters} \textbf{\bibinfo{volume}{88}},
  \bibinfo{pages}{033903} (\bibinfo{year}{2006}{\natexlab{a}}).

\bibitem[{\citenamefont{da~F~Costa et~al.}(2007)\citenamefont{da~F~Costa,
  Kaiser, and Hilgetag}}]{Costa2007b}
\bibinfo{author}{\bibfnamefont{L.}~\bibnamefont{da~F~Costa}},
  \bibinfo{author}{\bibfnamefont{M.}~\bibnamefont{Kaiser}}, \bibnamefont{and}
  \bibinfo{author}{\bibfnamefont{C.~C.} \bibnamefont{Hilgetag}},
  \bibinfo{journal}{BMC Syst Biol} \textbf{\bibinfo{volume}{1}},
  \bibinfo{pages}{16} (\bibinfo{year}{2007}),
  \urlprefix\url{http://dx.doi.org/10.1186/1752-0509-1-16}.

\bibitem[{\citenamefont{da~F.~Costa et~al.}(2007)\citenamefont{da~F.~Costa,
  Rodrigues, Travieso, and {Villas Boas}}}]{Costa2007}
\bibinfo{author}{\bibfnamefont{L.}~\bibnamefont{da~F.~Costa}},
  \bibinfo{author}{\bibfnamefont{F.~A.} \bibnamefont{Rodrigues}},
  \bibinfo{author}{\bibfnamefont{G.}~\bibnamefont{Travieso}}, \bibnamefont{and}
  \bibinfo{author}{\bibfnamefont{P.~R.} \bibnamefont{{Villas Boas}}},
  \bibinfo{journal}{Advances in Physics} \textbf{\bibinfo{volume}{56}},
  \bibinfo{pages}{167} (\bibinfo{year}{2007}).

\bibitem[{\citenamefont{da~F.~Costa
  et~al.}(2006{\natexlab{b}})\citenamefont{da~F.~Costa, Rocha, and {de
  Lima}}}]{Costa2006}
\bibinfo{author}{\bibfnamefont{L.}~\bibnamefont{da~F.~Costa}},
  \bibinfo{author}{\bibfnamefont{F.}~\bibnamefont{Rocha}}, \bibnamefont{and}
  \bibinfo{author}{\bibfnamefont{S.~M.~A.} \bibnamefont{{de Lima}}},
  \bibinfo{journal}{Physical Review E (Statistical, Nonlinear, and Soft Matter
  Physics)} \textbf{\bibinfo{volume}{73}}, \bibinfo{eid}{011913}
  (pages~\bibinfo{numpages}{10}) (\bibinfo{year}{2006}{\natexlab{b}}),
  \urlprefix\url{http://link.aps.org/abstract/PRE/v73/e011913}.

\bibitem[{\citenamefont{Fontoura~Costa and Silva}(November 2006)}]{Costa2006c}
\bibinfo{author}{\bibfnamefont{L.}~\bibnamefont{Fontoura~Costa}}
  \bibnamefont{and} \bibinfo{author}{\bibfnamefont{F.}~\bibnamefont{Silva}},
  \bibinfo{journal}{Journal of Statistical Physics}
  \textbf{\bibinfo{volume}{125}}, \bibinfo{pages}{841} (\bibinfo{year}{November
  2006}),
  \urlprefix\url{http://www.ingentaconnect.com/content/klu/joss/2006/00000125/%
00000004/00009130}.

\end{thebibliography}

\end{document}